\documentclass[fleqn,10pt]{wlscirep}

\usepackage{color}

\newcommand{\cut}[1]{{}}
\newcommand{\bigO}[1]{\ensuremath{\mathop{}\mathopen{}\mathcal{O}\mathopen{}\left(#1\right)}}

\title{Using complex networks towards information retrieval and diagnostics in multidimensional imaging}

\author[1]{Soumya Jyoti Banerjee}
\author[2]{Mohammad Azharuddin}
\author[3]{Debanjan Sen}
\author[3]{Smruti  Savale}
\author[3]{Himadri Datta}
\author[2]{Anjan Kr Dasgupta}
\author[1,*]{Soumen Roy}
\affil[1]{Bose Institute, 93/1 Acharya PC Roy Road, Kolkata 700 009, India}
\affil[2]{Department of Biochemistry, University of Calcutta, 35 Ballygunge Circular Road, Kolkata 700 019, India}
\affil[3]{Regional Institute of Ophthalmology, Calcutta Medical College and Hospital, Kolkata 700 073, India}

\affil[*]{soumen@jcbose.ac.in}


\begin{abstract}
We present a fresh and broad yet simple approach towards information retrieval in general and diagnostics in particular by applying the theory of complex networks on multidimensional, dynamic images. We  demonstrate a successful use of our method with the time series generated from  high content thermal imaging videos of patients suffering from the aqueous deficient dry eye (ADDE) disease. Remarkably, network analyses of thermal imaging time series of contact lens users and patients upon whom Laser-Assisted in situ Keratomileusis (Lasik) surgery has been conducted, exhibit pronounced similarity with  results obtained from ADDE patients. We also propose a general  framework for  the transformation of multidimensional images to networks for futuristic biometry.  Our approach is general and scalable to other fluctuation-based devices where  network parameters derived from fluctuations, act as effective discriminators and diagnostic markers. 
\end{abstract}

\begin{document}

\flushbottom
\maketitle
\thispagestyle{empty}

\section*{Introduction}

The field of Content Based Image Retrieval (CBIR) started with retrieval of specific images from a large array of images.   Nowadays, CBIR is more generally referred to as Content Based Multimedia Information Retrieval (CBMIR) or simply MIR. Information retrieval in general can be conceived of as finding material of an unstructured nature that satisfies an information need from within large collections~\cite{Raghavan-book}\@.  Applications of pictorial search into a database of images already existed in specialized fields like character recognition, face recognition, and, robotic guidance. IBM developed the first commercial CBIR system, called QBIC (Query by Image Content) in 1995~\cite{QBIC}\@. At present, the basic problem is the creation of powerful content-based methods in order to enable or improve multimedia retrieval\cite{ACMSIGMM}\@. Many interesting and challenging scenarios arise in case of real time videos, e.g. surveillance, live cell imaging in life sciences, biomedical imaging etc.

Multidimensional imaging (MDI) is a ubiquitous and integral part of modern life. It is used in extremely diverse fields ranging from entertainment or surveillance on one hand to science, medicine or surgery on the other. To thrive like any other successful technology, a typical MDI technique needs to be inexpensive, portable, robust and high resolution to the extent possible. Additionally, successful coupling with efficient  information retrieval algorithms could yield substantial premium. In this context, thermal imaging (TI) occupies an important place in MDI.  TI is an economical and potent yet relatively unexplored technique in medical diagnostics. Lower imaging resolution and variability of steady state thermal behavior due to environmental thermal fluctuations are seen as primary reasons for the restrictive use of this powerful non-invasive method. \\

\noindent 
{\bf  Graph Theory in Computer Vision: A Toplogical Perspective.}
Many problems in image processing can be naturally mapped to energy minimisation approaches. However, such energy minimisation problems could be highly demanding from the computational point of view, as the general requirement is to minimise a non-convex function in a space with thousands of dimensions. Thankfully, dynamic programming can be used, but, only in a limited number of cases, where the energy functions have special forms \cite{dynamic}\@. In absence of such privileges, researchers typically used global optimisation techniques like simulated annealing\cite{Geman}\@  or greedy algorithms \cite{Besag1986}\@ for image smoothing which would   be very slow for obvious reasons.

``Graph cut" approaches have come to be widely used in computer vision especially those that could be formulated in terms of energy minimisation. The essence of such approaches is that the basic technique is to construct a specialized graph on which the energy function to be minimized, such that the minimum cut on the graph in turn minimizes the energy. This follows from from the max-flow min-cut theorem that in a flow network, the amount of maximum flow is equal to capacity of the minimum cut.  It was shown  that  maximising the flow through an image network is associated with the maximum {\it a posteriori} estimate of a binary image, introduction of sources and sinks make the problem efficiently solvable \cite{Greig1989}\@.  These approaches have been used successfully in a wide variety of vision problems including shape matching\cite{matching}\@,  image restoration \cite{Boykov1 , Boykov2}\@, fingerprint recognition\cite{Fingerprint}\@,  surface fitting\cite{Surface}\@,  stereo and motion \cite{Boykov1 , Boykov2}\@ and medical imaging \cite{fMRI}\@. 

There also exists a body of work\cite{DAG-review, Applied_GT}\@  towards applying spectral encoding of a graph for indexing to large database of image features represented as Directed Acyclic Graphs (DAG). Databases of topological signatures can be indexed efficiently to retrieve model objects having similar topology. Significant research has been conducted on a general class of matching methods, called bipartite matching, to problems in object recognition The time complexity for finding such a matching in a weighted bipartite graph with $N$ vertices was determined as $\bigO {N^2 \sqrt{N log log N}} $\cite{Gabow}\@.

Recent researches on image segmentation have used multi-resolution community detection methods in fluorescent lifetime microscopy\cite{Nussinov-SPIE, Nussinov-JMI}\@. Replica inference approaches have also been used towards unsupervised multiscale image segmentation\cite{replica}\@.  Herein,  we have used graph theory from a different perspective. Instead of  object identification based on spatial correlations, we have exploited the relational topology of the image objects. This approach  adds another angle to image segmentation and object identification,  two classic problems in image processing.\\

\noindent 
{\bf  Time Series to Networks.} 
A large number of approaches to analyze time series have been proposed over time. These range from time-frequency methods, such as Fourier and wavelet transforms~\cite{Korner, Box, Percival}\@, to nonlinear methods, such as phase-space embeddings, Lyapunov exponents, correlation dimensions and entropies~\cite{Strogatz, Kantz, Campanharo}\@. These techniques are helpful for summarizing the characteristics of a time series into compact metrics. Such brevity can be efficiently exploited to effectively understand the dynamics or to predict how the system will evolve over time. However, these measures preserve many but not all of the important properties of a given time series. Therefore, there is considerable research toward the identification of metrics that can capture the additional information or quantify time series in a completely new ways~\cite{Zhang1, Lai, Verplancke, Ao}\@.  

Quite independent of the above, the field of complex networks has been extensively studied by itself and successfully applied in manifold instances in science, nature and engineering~\cite{Laszlo-rmp, Mark-book}\@. With significant advances being reported from various fields~\cite{pre091, Newman, RKG1, epl091, RKG2, interdependent, pre101, cri101, ssb121, RMP,  Girvan, pre151}\@, the importance of converting time series into networks is becoming increasingly clear over the last few years~\cite{Strozzi}\@.\\

\noindent 
{\bf From Videos to Time Series and Thence to Networks.} 
In this work, we furnish a new, simple and general route to information retrieval by combining  developments from these  disparate fields and show that such an approach can yield rich dividends for MDI in general and for diagnostics in particular. Indeed, following the broad framework proposed here, it is possible to construct inexpensive devices for non-invasive diagnostics and biometric based applications, which can  perform successfully in real-time~\cite{patent}\@.  Our method consists of the following steps: (i) conversion of a given video or MDI into time series, (ii) conversion of the time series into a network, and finally (iii) analysing the network metrics to identify specific  topological metric/(s) which can act as good discriminators for different videos.  \\

\noindent {\bf Advantages of the Present Approach.} The process of conversion of any given video to a time series has been known for some time\cite{time-series-patent}\@.  Albeit, to our knowledge, the fullest potential of this conversion in thermal imaging has not yet been  exploited. Effective utilization of the vast research in time series analysis and related advances is obviously critical to gain liberal advantage of this transformation in information retrieval.  

 A network based representation of time series, provides us with an analytical tool that may allow object identification, which is not possible in many conventional image processing techniques.  The uniqueness of the present identification approach is the use of analyses based on temporal instead of spatial distributions. As such network, based insights can be fed back for extraction of hidden image contents which are not evident from the spatial data alone.

\begin{figure*}[htbp]
\begin{center}
\includegraphics[width=5in]{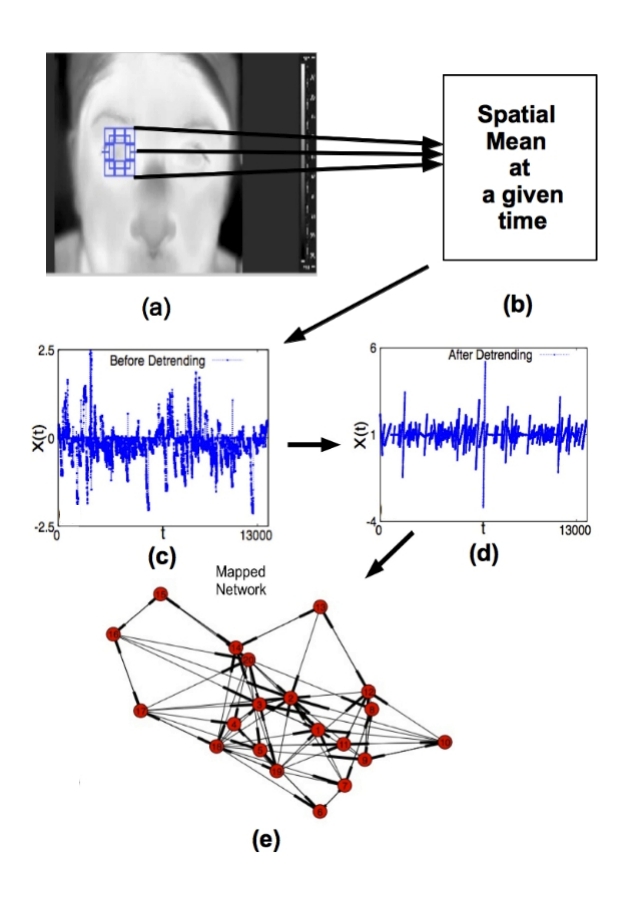}
\end{center}
\caption{{\bf Outline of our method.} (a) Thermal imaging of the right eye of a arbitrary individual, (b) spatial mean of the cropped region of interest at time $t$. Plot of pooled thermal time series data for both eyes of a particular group (like dry eye patients or healthy individuals) of all such thermal images (c) before detrending, and, (d) after detrending. Directed network obtained from mapping of detrended time series is shown in (e).}
\label{fig:our-methodology}
\end{figure*}

Principal component analysis or PCA \cite{Jolliffe}\@  is a potent and widely used linear transform in signal and image processing, more specifically in image compression, blind signal separation, face and pattern recogntion \cite{Gonzales, Barret, Petrou,Turk}\@ etc.  Essentially, PCA is a method for transforming a multidimensional dataset to a lower dimension.  The basis vectors follow modes of the greatest variance, when the  data is represented by PCA in the new coordinate system. However, PCA is also computationally expensive compared to many other processes like Fast Fourier Transformation.  Herein, we show  that in the present approach,  the computationally expensive procedure of PCA  adopted for dimensional reduction in conventional image processing can be safely circumvented.  Obviously, dimension reduction achieved by our method would make feature identification of complex videos and images computationally far simpler. As detailed below, our approach effectively opens up the avenue of fluctuation based diagnostics in biomedical MDI. Hardware implementations of this method is  extremely versatile, as it is smart, fast and portable\cite{patent}\@. 

No established method, to our knowledge, has addressed the problem of the dynamics of thermal behavior from source thermal imaging data. Conventional image processing algorithms typically  attempt appropriate segmentation, noise elimination and object identification like  morphological changes or relative pixel dynamics. For videos, images extensive work has been done on motion tracking and this known to have important implications in contexts like security and surveillance.

Lastly, The present work  of mapping biomedical videos into a time series and thence to a network  should  be implemented in diagnostic approaches, which need to record biomedical time-series data over a prolonged duration, perhaps without  rest. The inconvenience or pain caused to a patient is  imaginable.

\noindent {\bf Applications to Diagnostics: Dry Eye Disease.}  While the approach proposed in this paper  is very general,  herein,  we specifically concentrate on  patients with Aqueous Deficient Dry Eye (ADDE)  disease, contact lens users and patients who have undergone Lasik operations. We also investigate applications of our work in biometrics. ADDE is a disturbance in tear film physiology that leads to various abnormal states of ocular surface cells that elevate the incidence of ocular surface disorders and infection. ADDE represents one of the most common ocular pathologies and is a complex multifactorial disease characterized by an immune and inflammatory process that affects the lacrimal glands and ocular surface. Its diagnosis by  assessment of the tear film has been extensively studied~\cite{Tomlinson,Khanal,Goins}\@.  Most of the diagnostic approaches are based on either osmomolarity or evaporation of the tear film. Studies indicate that most ADDE measures do  not capture the etiologies for dry eye, such as dysfunctional neurology, hormonal influences or the inflammatory nature of the condition. Another situation  that may affect the alteration of the tear dynamics is the use of contact lens.  Statistical studies of thermal fluctuation of healthy individuals (control) and ADDE patients where non-invasive TI was used, have been conducted recently~\cite{Azharuddin}\@. Significant correlation of thermal fluctuations is found between left and right eye of control whereas this property is completely absent in eyes of ADDE patients.  However, the  problem of classification of dry eye either from collective or individual data remains unsolved. Also, parametric classification to differentiate or diagnose healthy and dry eye individuals is still unavailable.  The mechanism proposed here shows that thermal fluctuation based approaches and a robust parametrization of such fluctuation by network mapping may be a powerful alternative approach to express the etiology of the eye.  Throughout this paper, we use the terms dry eye and ADDE interchangeably. However, it should be especially noted that in medical literature, ADDE denotes only one of the spectra of alteration of ocular surfaces going by the name of the dry eye.

\section*{Methods}

\noindent {\bf Ethics statement.} All experiments analyzed herein were conducted after approval of the Ethics committee of Regional Institute of Opthalmology (RIO), Kolkata  and were carried out in accordance with the approved guidelines of RIO. The research adhered to the tenets of the Declaration of Helsinki of the World Medical Association. Informed consent was obtained from all subjects.\\

\noindent 
Fig.~\ref{fig:our-methodology} presents a schematic outline of our method.  Specific details of our method are  extensively discussed below. \\

\noindent
{\bf Thermal Imaging Setup.} For our experiments, we used a Forward Looking Infra Red (FLIR) thermal camera, Model no. FLIR SC 305, FLIR Systems AB, Sweden.  This camera is equipped with an RJ-45 gigabit Ethernet connection that supplies $16$ bit $320\times240$ images at rates as high as $60$ Hz along with linear temperature data. The video can be exported to several formats including AVI. In the FLIR SC 305 model, compression is used in the original video image and only the in built frame compression is used. Each frame is then cropped to select a region of interest (say eye, cheek etc). The camera was used with a thermal sensitivity of less than ${0.05}^{o}C$ at ${30}^{o}C$, spatial, temporal and image resolution of $1.36$ mrad, $9$ frames per second and $320\times240$ pixels respectively, with spectral range between 7.5 and 13 mm.\\

\noindent
{\bf Details of Data Collection and Clinical Background of Subjects.} 
Following are details of patient groups and healthy individuals for whom the thermal imaging videos were recorded for a duration of about $15$ second and subsequently analysed: 

\noindent
$\bf (a)$  $36$ Healthy individuals  or for $72$ eyes. Among them, $20$ were female and $16$ male, with a mean age of $28.4$ years. \\
$\bf (b)$  $42$ ADDE patients or for $84$ eyes.  Among them, $25$ were female and $17$  male, with a mean age of   $35.2$ years. \\
$\bf(c)$  $32$ patients who had Lasik surgery  or for $64$ eyes. Among them, $18$  were male and $14$  female, with a mean age of    $35.4$ years. \\
$\bf(d)$  $29$ Contact lens users or for $58$ eyes. Among them, $15$ were female and $14$ male, with a mean age of   $30.6$ years. \\
For (d), videos were separately acquired  for every individual when he or she was (i) wearing  lens, and, (ii) not wearing  lens.

The ocular surface temperature were recorded with eyes open and the subjects were asked not to blink during the recording. Noise of individual data could come from blinking of eyes if the videos are recorded for a longer duration. Probability of blinking of eyes tends to zero in  a small duration like $15$ second and therefore noise is negligible for the recorded  data.\\

\noindent {\bf Conversion of Videos to Time Series: PCA or a simple arithmetic mean?} 
 Frame wise conversion of TI videos to time series has been implemented two ways: (1) Principal component analysis (PCA) of the cropped regions at each time instant, $t$, and, (2) Mean pixel value of the cropped region at each time instant, $t$.  Thus, two entirely different time series of $N$ points are obtained by the above methods. As is well-known,  PCA is a frequently used technique for dimensional reduction in image processing~\cite{Gonzales, Barret, Petrou,Turk}\@.  As observed in Fig.~ S 1 of  supplementary information (SI),  there is no significant variation of the principal components over time.  The first three principal components collectively retain about $82\%$ of variance of the data, in which the first principal component alone accounts for about $55\%$.  
On comparison between  these two time series, we find that the classification of data, achieved by using mean pixel value as witnessed in Fig. \ref{fig:control-dry}, performs as good as the leading Principal Component, which is demonstrated in Fig.~ S 2 of SI.  The first three principal components capture about $82\%$ of variance of the data, in which the first principal component alone accounts for over $55\%$.  Naturally, we prefer the mean pixel value over PCA because, it:  (i) is computationally far less inexpensive, (ii) is easy to implement in hardware and can be used to construct fast, smart and portable devices \cite{patent}\@, and, (iii) provides a statistically  unambiguous and robust classification measure for a wide variety of circumstances. The latter is clearly evident in Fig.~\ref{fig:control-dry}  to Fig.~\ref{fig:Lasik}. Thus we have taken the average pixel value as instantaneous temperature of the selected region, $\tau(t)$, at time instant, $t$. The initial instantaneous temperature, $\tau(0)$, at $t = 0$, is then set to $0$. Thence, we obtain our final  time series, $X(t) = \tau (t) -\tau (0)$.\\

\noindent {\bf Detrending and Pooling Time Series.}
Some linear trend from decay of eye temperature is likely to be present in the time series generated from every video due to evaporation of water from the eye. In the event, that a large single time series is generated by pooling each of these smaller time series,  a pseudo-periodicity could be created, which could mislead our predictions. So we detrend each time series separately to strip the accumulated time series from such a linear trend and normalize it by mean of all detrended values to fixed amplitude base line as shown in {Figs.~\ref{fig:our-methodology}(c) and \ref{fig:our-methodology}(d).

We independently pool the time series data of (1) both left and right eyes collectively, and, (2) the cheek of every individual in a given eye group (control or dry)  and create a single time series. In all,  we have eight classes of pooled time series. Throughout the rest of the paper, we often use $\{{\cal H_E}\},  \{\cal D_E\}, \{\cal H_C\}, \{\cal D_C\}, \{\cal C_L\}, \{\cal C_N\}, \{\cal L_B\}$ and $\{\cal L_A\}$ to denote: (i) healthy eyes (control group), (ii)  dry eye group, (iii) cheek (control group), (iv) cheek (dry eye group), (v) group of contact lens users wearing lens, (vi) group of contact lens users not wearing lens, (vii) group of patients who underwent Lasik operations - before the procedure, and,  (viii) group of patients who underwent Lasik operations - after the procedure. It should be noted that the individuals in  (i) $\{\cal C_L\}$ and $\{\cal C_N\}$, and, (ii) $\{\cal L_B\}$ and $\{\cal L_A\}$ are identical.\\

\noindent {\bf Creation of Networks from Time Series: Background.}
In literature, there are a few interesting approaches for time series to network mapping, based on different concepts such as correlations~\cite{Zhang2, Yang}\@, visibility~\cite{Lacasa, Luque}\@, recurrence analysis~\cite{Marwan}\@, transition probabilities~\cite{Nicolis, Li, Shirazi}\@ and phase-space reconstructions~\cite{Xu, Gao}\@. A complete list of all the proposed maps can be found in Donner {\it et al.}~\cite{Donner}\@ and references therein. These studies are able to capture that distinct properties of a time series can be mapped onto networks with distinct topological properties. These findings claim that it may be possible to differentiate different time series features using network metrics. But it was unclear as to how these network topological properties are related to the original time series.  The main drawback of these maps $M : T  \rightarrow G$ from the time series domain $T$ to the network domain $G$ however, is that they do not have a natural inverse mapping $M^{-1} : G \rightarrow T$ . Recently, some researchers have tried to construct an inverse map~\cite{Strozzi, Shirazi, Haraguchi}\@. Nonetheless, these methods are either sensitive to arbitrarily chosen parameters~\cite{Strozzi, Haraguchi}\@ or are dependent on the information about the given map $M$ for construction of an inverse mapping $M ^ {-1}$~\cite{Shirazi}\@. Therefore, they are not immensely useful for real world networks, when $M$ is not known in advance.Herein, we have used a recently developed time series to network mapping technique~\cite{Amaral}\@. In our work, we have used only the forward mapping from time series to directed networks.\\

\noindent {\bf Salient Features of Mapping Time Series to Directed Networks.} 
We have followed  methods well-established in  literature~\cite{Amaral}\@, for mapping the detrended time series of each of the eight groups mentioned above into eight distinct directed networks. These directed networks are denoted by ${\cal G}_i ({\cal V}_i, {\cal E}_i)$, where, ${\cal V}_i$ and ${\cal E}_i$ is the set of all nodes and edges respectively in network, and,  $i \in\{ {\{\cal H_E}\}, \{\cal D_E\}, \{\cal H_C\}, \{\cal D_C\}, \{\cal C_L\}, \{\cal C_N\}, \{\cal L_B\}, \{\cal L_A\} \}$. Further, for our case, we divide the Y-axis of Fig.~\ref{fig:our-methodology}(d) into $20$ quantiles.  It has been shown earlier \cite{Amaral}\@, that the  time series to network mapping method used herein, is broadly independent of the number of equiprobable quantiles. Furthermore, the number of quantiles has no significant effect over the network topology\cite{Amaral}\@. We have chosen $20$ quantiles to create a network with reasonable number of nodes and edges. We believe this to be an optimal number which can be accepted without the loss of generality, for reasons explained below. Creating too large a network will lead to unnecessary increase of computation time for calculating values of all  node based and edge based network metrics, which are integral to our approach. Again, a very small number, e.g. $5$ quantiles, is an unreliable choice for classification of groups using network properties.  In our analyses, we have {\em ignored edge weights} because we are strongly  interested {\em in capturing important thermal fluctuation transitions from one quantile to another rather than in how many times such transitions occurred.}

Nodes and edges of the networks mapped from the thermal videos have a clear physical interpretation. The nodes of the directed networks signifiy different ranges of temperature and edges represent the transitions from one temperature range to another. These obviously could be from a lower temperature regime to a higher temperature regime or vice versa. But, mere visual examination of networks derived by mapping from pooled thermal imaging time series is generally insufficient to classify between  healthy eye and dry eye groups,  irrespective of the network layout, as shown in Fig. S 3 and Fig. S 4 of Supplementary Information.  Therefore, clearly identifying distinguishing properties of the thermal fluctuations for healthy individuals and diseased patients will be possible only upon thorough analyses of the directed networks  using different topological network metrics and not merely by visual examination. \\

\noindent {\bf Analyzing Directed Networks.}
All  directed networks, thus obtained by mapping from thermal time series have been analyzed thoroughly using most known network metrics~\cite{epl091}\@, as  detailed in Fig. S 5 to Fig. S 8 of Supplementary Information. As shown therein, among all network metrics, we find that edge betweenness centrality~\cite{Newman, Girvan}\@, ${\cal B}_e$, demonstrates significant discriminating power, a fact that can well be exploited to construct a functional, scalable device~\cite{patent}\@. For directed networks, ${\cal B}_e$, $e\in{\cal E}_i$ is the ratio of the number of directed shortest paths, {$\sigma {(s, t | e)}$, between node, $s$, and node, $t$, which pass through edge, $e$, and the total number of directed shortest paths, {$\sigma {(s, t)}$, between node, $s$, and node, $t$, in the network. Mathematically, it is defined as:
\begin{equation} 
{\cal B}_e = \displaystyle\sum_{s \not= t} \frac{\sigma (s, t | e)}{\sigma (s, t)}
\label{betweenness}
\end{equation}
We then construct the {\em cumulative distribution} of ${\cal B}_e$,  for each of the above groups.

\section*{Results}

\noindent{\bf Cumulative Edge Betweenness Centrality Distribution for Dry Eye Versus Healthy Eye.}  
From the cumulative ${\cal B}_e$ distributions shown in Fig.~\ref{fig:control-dry}  it is clear that  the  networks mapped from pooled TI time series of eye for healthy eyes (control) group, $\{{\cal H_E}\}$,  possess lower  ${\cal B}_e$ values where as the  networks mapped from pooled TI time series of eye for dry eyes group, $\{{\cal D_E}\}$,  possess higher ${\cal B}_e$ values.  Since dry eye is a pathology or the eye; from  Fig.~\ref{fig:control-dry} we get a hint that presence of edges with high  ${\cal B}_e$  value could possibly be a suitable distinguishing, diagnostic feature for multiple pathological conditions in eyes. We subsequently find that this is indeed true to a large measure.\\

\begin{figure}[htbp]
\begin{center}
\includegraphics[width=3in]{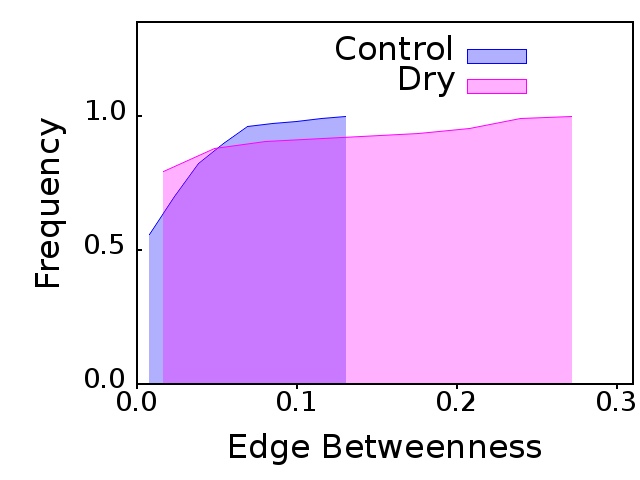}
\caption{{\bf Healthy eye versus dry eye.} Cumulative distribution of edge betweenness centrality for networks mapped from pooled thermal imaging time series for healthy eye and dry eye groups.}
\label{fig:control-dry}
\end{center}
\end{figure}

\noindent
{\bf Diagnosing Patients with Dry Eye.}
As aforementioned thermal imaging videos of $15$ second duration were obtained for $42$ dry eye patients, $36$ healthy individuals, $32$ individuals who underwent Lasik surgery, and, $29$ regular contact lens users. For the Lasik Group, videos were obtained for all patients both before, and, after the surgery. For Contact lens users, videos were obtained both when they were: (a) wearing their lenses, and, (b) not wearing their lenses. In addition, TI videos of healthy individuals and patients with dry eye were further recorded for independent validation. These ``test case" videos were of a total duration of $1$ minute and obtained by pooling TI Videos of $15$ second duration of the same individual. Such pooling is necessary, because: (i) blinking of eyes is not allowed in our experiment and it is difficult for most individuals not to blink even once in a minute, and, (ii) proper rest needs to be given to the eyes of all participants in the experiment.

\begin{figure}[htbp]
\begin{center}
\includegraphics[width=3in]{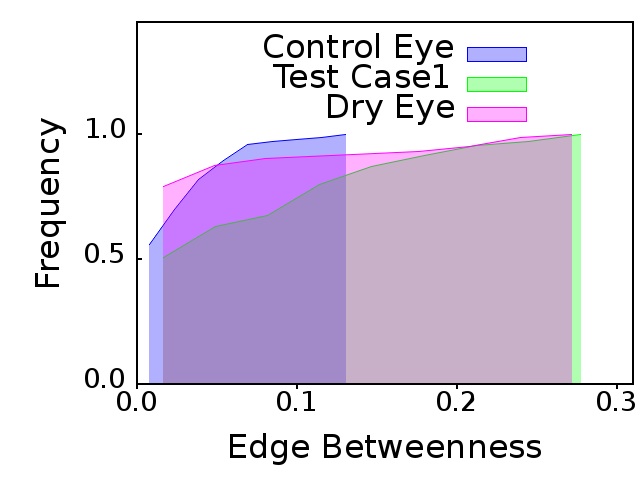}
\end{center}
\caption{{\bf Test case: dry eye patient.} Cumulative distribution of edge betweenness centrality, ${\cal B}_e$, for networks mapped from pooled thermal imaging time series for  individuals with healthy eyes (control) and dry eye patients shown in Fig.~\ref{fig:control-dry} and a test case. The test case is that of a dry eye patient whose cumulative ${\cal B}_e$ distribution is almost identical to the cumulative ${\cal B}_e$ distribution of networks mapped from pooled thermal imaging time series data for all other dry eye patients.}
\label{fig:control-dry-test1}
\end{figure}

From Figs.~\ref{fig:control-dry}, \ref{fig:control-dry-test1}, \ref{fig:control-dry-test2}, \ref{fig:dry-eye-cheek} and \ref{fig:control-eye-cheek},  it is  clear that the value of ${\cal B}_e = 0.2$  in the distribution of ${\cal B}_e$,   serves as a cut-off  towards clearly distinguishing between healthy individuals and ADDE. If the cumulative ${\cal B}_e = 0.2$ distribution of a subject, whose eyes have been thermally videographed for a minute, with adequate intermittent resting  crosses  ${\cal B}_e = 0.2$, the subject is an ADDE patient. On the other hand, the cumulative ${\cal B}_e$ distribution of a  healthy subject who has been similarly videographed, should  clearly fall short of  ${\cal B}_e = 0.2$.  Thus, the present method has an inbuilt potential for incorporating a distance-like measure,  ${\cal B}_e$, for clear image  based segregation. It should be noted that the cutoffs seem statistically robust.  Figs.~\ref{fig:dry-eye-cheek} and \ref{fig:control-eye-cheek} have each been plotted for  $15$ pairs of randomly selected eyes from the original dataset of  $42$ pairs of dry eyes, and, $36$ pairs of healthy eyes. The cutoffs remain unaltered from that derived in Fig.~\ref{fig:control-dry} for the original dataset. \\

\begin{figure}[htbp]
\begin{center}
\includegraphics[width=3in]{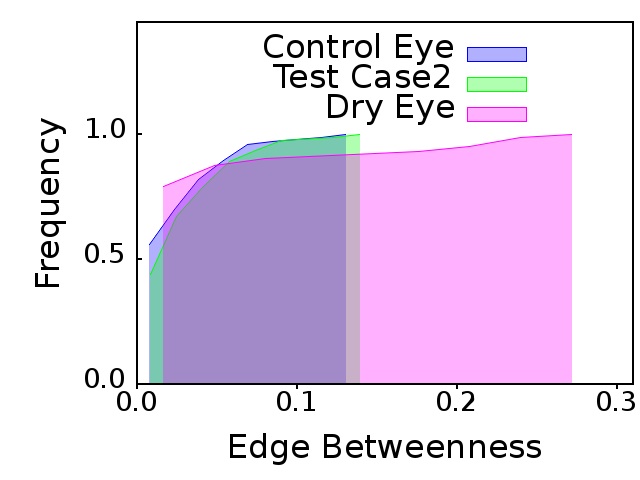}
\end{center}
\caption{{\bf Test case: individual with healthy eye.} Cumulative distribution of edge betweenness centrality, ${\cal B}_e$, for networks mapped from pooled thermal imaging time series for individuals with healthy eyes (control) and dry eye patients shown in Fig.~\ref{fig:control-dry} and a test case. The test case is that of a healthy individual whose cumulative ${\cal B}_e$ distribution is almost identical to the cumulative ${\cal B}_e$ distribution of networks mapped from pooled thermal imaging time series data for all healthy individuals.}
\label{fig:control-dry-test2}
\end{figure}

\noindent
{\bf{Adoption of a Double Blind Procedure.}}
The following three completely disjoint groups have conducted this work:  
$(i)$ opthalmologists {\em (Group A)},  
$(ii)$ experimentalists who undertook thermal imaging {\em (Group B)}, and, 
$(iii)$ theoretical analysts, who had no exposure to the patients and were not present during videography {\em (Group C}). Thus, our diagnostic procedure is double blind and  was conducted in the following  steps.
\begin{itemize}
\item{{\em Step 1:} {\em Group A} performed standard ADDE diagnostic tests upon classes of healthy individuals and patients but identified them only as belonging to either $Class~1$ or $Class~2$ to {\em Group B}; thus blanking out any clinical information. Additionally, two individuals of different clinical background, who were willing to be videographed for a longer duration were also supplied to {\em Group B}, again without any information about their clinical background.} 
\item{{\em Step 2:} {\em Group B} undertook non-invasive thermal imaging without any knowledge of the clinical background of their subjects and shared all videos of $Class~1$ and $Class~2$ and also of the two individuals to {\em Group C}.} 
\item{{\em Step 3:} {\em Group C} theoretically analysed the videos as detailed in this section and arrived at the cutoff of ${\cal B}_e = 0.2$. With help of this cutoff predicted by the training set data, namely, $Class~1$ and $Class~2$;  this group could predict whether the two test case individuals belonged to $Class~1$ or to $Class~2$.} 
\item{{\em Step 4:} Lastly {\em Group A} confirmed that the inference of {\em  Group C} is correct, i.e. Test case $1$ is actually an ADDE patient, and, Test Case $2$ is definitely a healthy individual, as predicted in Figs.~\ref{fig:control-dry-test1} and \ref{fig:control-dry-test2} respectively.  Notably, these two test case eyes do not belong to any of the classes of the training sets, namely,  $Class~1$ and $Class~2$.} 
\end{itemize}

\begin{figure}[htbp]
\begin{center}
\includegraphics[width=3in]{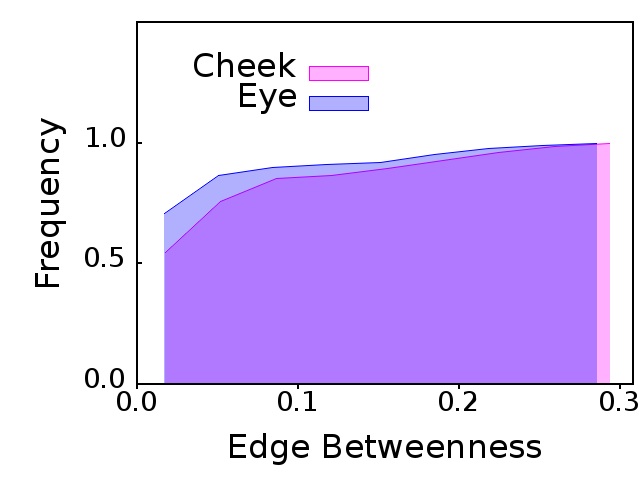}
\end{center}
\caption{{\bf Cheek versus eyes of dry eye patients.} Cumulative distributions of edge betweenness centrality, ${\cal B}_e$ of networks mapped from pooled thermal imaging time series for $15$ individuals with dry eyes, and, their cheeks.}
\label{fig:dry-eye-cheek}
\end{figure}

From the cumulative ${\cal B}_e$ distributions shown in Fig.~\ref{fig:control-dry-test1},  we observe that cumulative ${\cal B}_e$ distribution of Test case 1 eye individual has larger shift toward right. The network mapped from TI time series for  test case1 eye network has larger number of edges with high ${\cal B}_e$ values which is quite similar to dry eye group.  On the other hand, in the network obtained for the Test case 2 eye individual, the  ${\cal B}_e$ distribution is akin to that of the control eyes group as seen in Fig.~\ref{fig:control-dry-test2}.  Therefore, we can reasonably infer that test case 1 should be a dry eye and test case 2 should be a healthy eye. Thus, in practice, diagnosis of dry eye is indeed possible by non-invasive thermal imaging of the eye of the patient for $60$ second,  by sufficiently resting the eyes of the arbitrary individual after every $15$ second.\\

\begin{figure}[htbp]
\begin{center}
\includegraphics[width=3in]{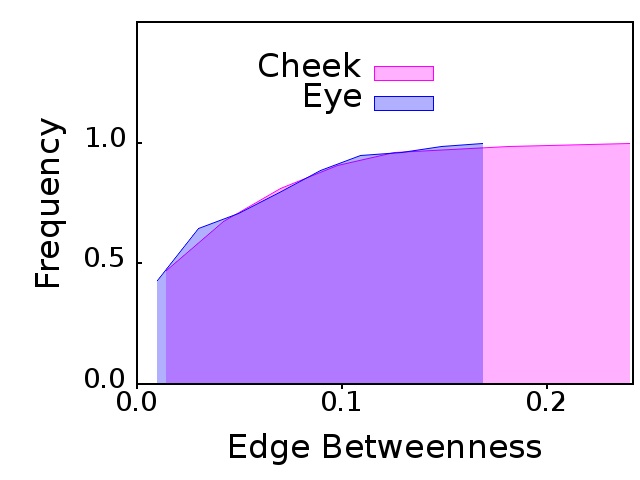}
\end{center}
\caption{{\bf Cheek versus eyes of healthy individuals.} Cumulative distributions of edge betweenness centrality, ${\cal B}_e$ of networks mapped from pooled thermal imaging time series for $15$ individuals with healthy eyes, and, their cheeks.}
\label{fig:control-eye-cheek}
\end{figure}


\noindent
{\bf Tagging Objects in Images.}
To identify possible biometric applications from our method, we also investigated the thermal fluctuations of other portions of the face apart from the eyes, namely, the cheek. The very  same facial videos of both control eye and dry eye groups were studied for this purpose. This time however, the cheek portion was selected. $15$ identical  individuals were randomly selected from $\{\cal H_E\}$ and $\{\cal H_C\}$. Similarly, $15$ identical  individuals were also randomly selected from $\{\cal D_E\}$ and $\{\cal D_C\}$. We {\em independently} pooled $15$ detrended thermal times series, obtained from cheek and eye TI time series data for $\{\cal H_E\}$, $\{\cal H_C\}$, $\{\cal D_E\}$ and $\{\cal D_C\}$.

Very interestingly, the cumulative edge betweenness, ${\cal B}_e$, distributions for eye and cheek show completely different behavior.  In the former case, the cumulative ${\cal B}_e$  distribution of individuals with dry eyes selected from $\{\cal D_E\}$ and $\{\cal D_C\}$ are almost identical, as can be observed from Fig.~\ref{fig:dry-eye-cheek}.  However, in the latter case, cumulative ${\cal B}_e$  distribution of individuals with healthy eyes selected from $\{\cal H_E\}$ and $\{\cal H_C\}$ are very different, as visible in Fig.~\ref{fig:control-eye-cheek}.

The above observations seem extremely important from the  point of view of biometrics, for the following reasons. If there is any metabolic or functional object in which a biochemical or chemical activity is in progress, the temporal data transformed in network can provide a parametric identification of the status of the process, and the regional dependence of the same can further provide information regarding aberrations in functional status if any.\\

\begin{figure}[htbp]
\begin{center}
\includegraphics[width=3in]{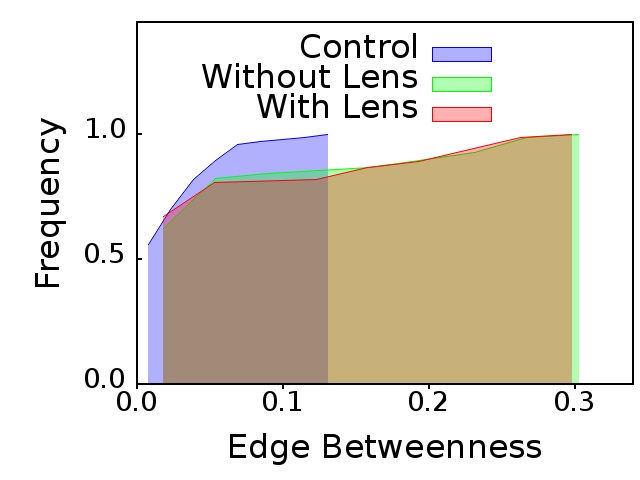}
\end{center}
\caption{{\bf Contact lens users.} Cumulative distributions of edge betweenness centrality, ${\cal B}_e$ for networks mapped from pooled thermal imaging time series for  individuals with healthy eyes, and, contact lens users. For the latter category, thermal imaging was conducted when they were: (a) wearing their contact lens, and, (b) not wearing their contact lens.}
\label{fig:control-lens}
\end{figure}

\begin{figure}[htbp]
\begin{center}
\includegraphics[width=3in]{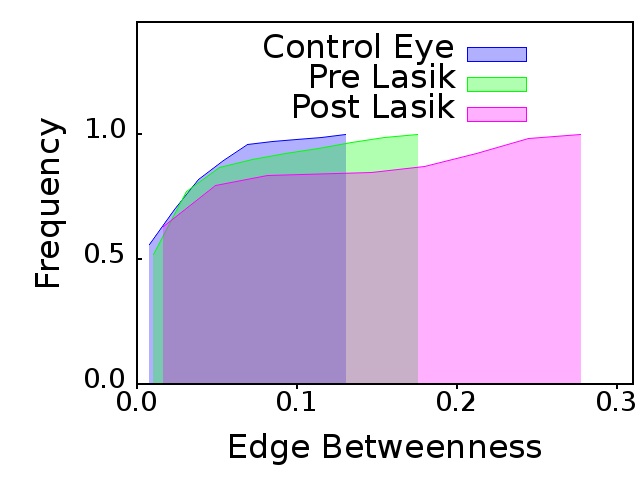}
\end{center}
\caption{{\bf Patients who underwent Lasik surgery.} Cumulative distributions of edge betweenness centrality, ${\cal B}_e$ for networks mapped from pooled thermal imaging time series for  individuals with healthy eyes  and eyes of patients upon whom Lasik operation was conducted. Thermal imaging was conducted on the latter category, both, before they underwent surgery, and, after they underwent surgery.}
\label{fig:Lasik}
\end{figure}


\noindent
{\bf Contact Lens Users.}
From the cumulative distributions shown in Fig.~\ref{fig:control-lens}, it is clear that the contact lens users, (a) who are wearing lens, $\{\cal C_L\}$, and, (b) who are not wearing lens, $\{\cal C_N\}$; show characteristics similar to the dry eye group, $\{\cal D_E\}$, i.e., presence of edges with high ${\cal B}_e$ values when compared to the group of healthy individuals, $\{\cal H_E\}$. Contact lens users  obviously also have unhealthy eyes due to myopia or hypermetropia. Significantly, removal of contact lens has almost no effect on the nature of the  distribution.\\

\noindent
{\bf Patients Undergoing Lasik Surgery}
From the cumulative ${\cal B}_e$  distributions shown in Fig.~\ref{fig:Lasik}, we observe that cumulative ${\cal B}_e$ distribution of patients who have undergone Lasik surgery is also very similar to that of the cumulative ${\cal B}_e$ distribution of dry eye group; just like the case of contact lens users.\\


\section*{Discussion}

Edges of the directed networks represent the transitions from one temperature range to another; for example from a lower temperature regime to higher temperature regime or vice versa. For this reason, edge based network metrics would be very suitable for classification of thermal time series. Thus, it is not a surprise that edge betweenness centrality, ${\cal B}_e$, captures the most important transitions, representing significant thermal fluctuations, between two completely different temperature regimes, and, in the process exhibits such clear discriminating properties. 

The results presented here illustrate that network based high content imaging can be a powerful tool for classification of objects in an image. The discriminatory potential of the said approach, which is in full display for ADDE patients in Fig. \ref{fig:control-dry},  contact lens users in Fig. \ref{fig:control-lens}, and, Lasik patients in Fig. \ref{fig:Lasik}; can hardly be derived using conventional video processing techniques.

As aforementioned,  to our knowledge no established method has addressed the problem of the dynamics of thermal behavior from source thermal imaging data.  In the present case, the image dynamics  is a variation of another function, namely,  temperature and the fluctuations are not due to environmental tuning alone as evident from above results. The function is actually a coupled manifestation of the thermal fluctuations of the environment and metabolic fluctuations.  This  coupling of fluctuations is difficult  to understand using conventional techniques alone.  
The network metrics  clearly imply that the coupling is different for healthy individuals and dry eye patients.  In healthy individuals, the fluctuations in the eye remain independent of the other facial areas more exposed to the environment.  In dry eye patients, this coupling is much stronger and the fluctuations of the eye and cheek are similar,  implying that the metabolic control is lessened in the diseased state.  Thus, dry eye subjects show significantly more fluctuations as compared to healthy individuals.

Our future plan is to design a method where parallel dynamics of different regions can be simultaneously evaluated. This will  combine image processing algorithms like automatic object specific cropping and segmentation with  topological properties of networks. 
While the immediate diagnostic potential of our findings for contact lens users or Lasik operated patients is unclear, the remarkable similarity of the cumulative ${\cal B}_e$ distribution for all eye diseases investigated here, namely, ADDE, contact lens users who obviously suffer from myopia or hypermetropia and Lasik operated patients is indeed striking. The last finding is specially interesting in context of the ongoing debate in ocular medicine as to whether Lasik operation actually leads to deterioration in the condition of the patients eye.

Last and certainly not the least, it may be noted that  network based imaging approach can be exploited for designing  efficient alert systems  or  biometric codes.  Similarly, the technique can be applied to other forms of non-thermal high content imaging. The temporal dynamics  provides us with a toolbox that would serve to compliment the  knowledge gained from conventional image processing.


 \section*{Acknowledgments}
SJB and MA thank Council of Scientific and Industrial research, India and University Grants Commission, India respectively for financial support. 

 \section*{Author Contributions} 
SR and ADG designed research. SR wrote the manuscript. SJB  conducted theoretical analysis. MA conducted the thermal imaging experiments.  DS, SS and HD provided clinical supervision. 

\section*{Competing financial interests} 
The authors declare no competing financial interests.

\end{document}